# A trilinear quantum dot architecture for semiconductor spin qubits

R. Li*[1], V. Levajac[1,3], C. Godfrin[1], S. Kubicek[1], G. Simion[1], B. Raes[1], S. Beyne[1], I. Fattal[1,2], A. Loenders[1,2], W. De Roeck[3], M. Mongillo[1], D. Wan[1], K. De Greve[1,2]

[1] IMEC, Leuven, Belgium
[2] Department of Electrical Engineering, KU Leuven, Leuven, Belgium
[3] Department of Physics, KU Leuven, Leuven, Belgium

## Abstract

Semiconductor quantum dot spin qubits hold significant potential for scaling to millions of qubits for practical quantum computing applications, as their structure highly resembles the structure of conventional transistors. Since classical semiconductor manufacturing technology has reached an unprecedented level of maturity, reliably mass-producing CMOS chips with hundreds of billions or even trillions of components, conventional wisdom dictates that leveraging CMOS technologies for quantum dot qubits can result in upscaled quantum processors with thousands or even millions of interconnected qubits. However, the interconnect requirements for quantum circuits are very different from those for classical circuits, where for each qubit individual control and readout wiring could be needed. Although significant developments have been demonstrated on small scale spin qubit systems, qubit numbers remain limited, to a large extent due to the lack of scalable qubit interconnect schemes. Here, we present a trilinear quantum dot array that is simple in physical layout while allowing individual wiring to each quantum dot. By means of electron shuttling, the trilinear qubit architecture provides qubit connectivity that is equivalent to or even surpasses that of 2D square lattice qubit arrays. Assuming the current qubit fidelities of small-scale qubit devices can be extrapolated to large-scale arrays, even medium-length shuttling arrays on the order of tens of microns in length would already allow the realization of million-scale qubit systems, while maintaining manageable overheads. Beyond the qubit architecture, we also present a scalable control scheme, where the qubit chip is 3D-integrated with a low-power switch-based cryoCMOS circuit for parallel qubit operation with limited control inputs. As our trilinear quantum dot array is fully compatible with existing semiconductor technologies, this qubit architecture represents one possible framework for future research and development of large-scale spin qubit systems.

## Introduction

Fault tolerant quantum computers and practical quantum computation algorithms require the number of physical qubits to be in the million scale[1]. Controlling and interfacing such large quantum processors is a giant system-level challenge. A quantum computer is therefore expected to require a multi-layer architecture[2,3], a hierarchical structure not that dissimilar from classical computers. Along with this analogy, and in view of the complexity needed, both qubits and the lowest level control machinery of a quantum computer would benefit a lot if the unprecedented sophistication of classical semiconductor technology could be leveraged. This technology, and particularly CMOS processing, enabled scaling of the number of transistors on a single integrated chip from millions back in the 1990s to hundreds of billions nowadays[4].

Among different qubit technologies, semiconductor quantum dot spin qubits most closely resemble CMOS transistor technology[5–8]. From a qubit perspective, operation fidelities beyond the quantum error correction threshold have been demonstrated[9–16] in small scale quantum dot spin qubit systems in academic laboratories, and proof-of-principle long-range interactions have also been demonstrated – providing an existence proof for upscaling considerations[17–24]. Recently, also quantum dot spin qubits fabricated by fully CMOS-foundry-compatible methods have begun to show high device yield, good uniformity, and performance comparable to or even beyond those of the best academic-lab results[25–27]. Despite this progress, a clear roadmap to upscaling quantum dot qubits has been found lacking: in spite of extensive efforts, results on quantum dot spin qubit arrays have so far been limited to systems with qubit numbers on the order of 10, far below the desired million-scales[28–31]. These limitations can be partially explained by the small size of quantum dot spin qubits, which has the effect of a double edged sword on qubit scaling – while the small size allows for high density integration, the wiring interconnection at very tight and sustained pitches is challenging and quickly becomes the bottleneck[8,32]. This is additionally complicated by the fact that - unlike transistor circuits, where inputs and outputs can be cascaded – each qubit requires individual control and readout. Addressing those scaling challenges is not straightforward, as wiring fanout and crosstalk correction favor simple one-dimensional qubit arrays, while quantum algorithms and error correction rely on more complex connectivity that is inherent to two-dimensional qubit grids[33,34] or beyond. There are many studies trying to address the wiring challenges from different aspects[8,35–41]. However, one or more elements in those proposed architectures generally require technologies well beyond the current state-of-the-art CMOS fabrication capabilities, such as compact co-integration of qubits with on-site control circuits, or high qubit uniformity at the spin qubit level[35,36]. One solution could be separating qubits (or qubit arrays) with long range qubit couplers to open spaces for wiring interconnects[8,37–39]. However, long range couplers, either shuttling or spin-photon coupling based, also require a certain level of uniformity, and the quantum error correction schemes need further developments to consider the reduced connectivity of these couplers. For the above reasons, no complete error correction scheme has been implemented on quantum dot spin qubit systems so far.

Here, we present a linear spin qubit architecture that allows realistic individual qubit wiring, while having a qubit connectivity equivalent to that of a 2D lattice grid. The core of our proposed architecture is based on earlier work, where we and others proposed how a 2D qubit lattice grid could be mapped onto a bilinear qubit array for both superconducting and silicon spin qubits, with overlapping resonators providing the qubit connectivity between the two linear qubit arrays[42,43]. However, for silicon quantum dot spin qubits, resonator-mediated two qubit gates tend to be challenging due to the inherently limited coupling strength, resulting in limited fidelities thus far[20]. Also, the required dense arrays of resonators with well-matched qubit frequencies impose multiple, stringent engineering challenges. In this work, we therefore propose introducing instead a third linear array of quantum dots, in-between the two linear qubit arrays, to enable a similar qubit connectivity – but this time enabled via qubit shuttling. Unlike conventional shuttling-based long-range connections that rely on shared gate control to free up the space for interconnects, this trilinear architecture allows for connecting each individual quantum dot in the shuttling array and applying different voltages separately, in turn allowing tuning each in the array for best shuttling performance[18]. We discuss the physical implementation of our trilinear architecture for different system sizes, while also addressing the wiring interconnect bottleneck with 3D integration and cryoCMOS. We further discuss how to bypass potentially malfunctioning quantum dot sites in the proposed architecture, and how to achieve even higher order qubit connectivity - beyond the nearest-neighbor interactions of conventional 2D lattice grids.

## Trilinear quantum dot array

Figure 1 shows the concept of the mapping of a 2D lattice grid onto a trilinear array. For illustration purposes, we use a 4x4 lattice as shown in Fig. 1a. Each spin qubit is defined by a quantum dot, as denoted by the filled blue circles, and is connected to its nearest neighbors, as denoted by the blue lines. Fig. 1b shows the trilinear array as a direct mapping of the 2D lattice, where each odd (even) row from the 2D lattice is placed into the upper (lower) 1D array[43]. Two-qubit interactions within the row direction (green line in Fig. 1a) remain in the trilinear array and can be performed directly, as represented by the green line between the corresponding sites in Fig. 1b. For two-qubit interactions along the column direction in the 2D lattice (purple line in Fig. 1a), interactions between the upper and lower 1D array are required in the trilinear array. We implement such an operation by shuttling one qubit from the lower or upper 1D array to the middle array, which is composed of empty quantum dots - as shown by the empty blue circles in Fig. 1b. Then the qubit is shuttled along the middle array to the desired site (still in the middle array, next to the other qubit) to perform the two-qubit operation and shuttle back. This operation procedure is represented by the purple line in Fig. 1b.

The trilinear architecture could simplify both the physical qubit structure as well as their operation schemes[26,40,41,44–46]. The physical structure can be straightforwardly implemented with current CMOS technology as the gate pitch and critical dimensions are in the few tens of nanometer range – a range currently achieved for small-scale qubit demonstrators[26,47,48]. As shown in Fig. 1c, two wiring layers (blue and green) would suffice for fanout, allowing full control of the chemical potential of the quantum dots (via the circular plunger gate) and their coupling to the neighbouring dots (via the rectangular barrier gate) for each quantum dot. For simplicity of the discussion, here we omit the control and readout periphery which can be straightforwardly added to the scheme (see Supplementary Information S1). A key advantage of the linear nature of the proposed architecture layout is that such and other (linearly scaling) physical structures can be readily added to the layout. For example, an array of nanomagnets can be fabricated on top of the quantum dots for qubit addressability, and an array of single electron transistors can be added on each side along the trilinear array for the qubit readout. These could be implemented in the same gate layer as the quantum dot gates and are readily available with the current CMOS technology[45,49]. The separate wiring enables individual qubit fidelity optimization for every step of the quantum operations, including single and two-qubit gates and shuttling. Additionally, the device tune-up and crosstalk compensation will be more straightforward here than in 2D lattices[29,50–52]. Moreover, individual control also enables parallel operations at different qubit sites, and multiple qubits can be shuttled simultaneously in the middle quantum dot array[45].

In comparison to a 2D lattice, the trilinear qubit structure requires extra shuttling for two-qubit operations along the column directions. As shown in Fig. 1b, the adjacent 2D odd and even rows are at the upper and lower 1D qubit array, respectively, with one shifted horizontally by half of the row size. This means that the shuttling length in the middle array is half of the row size. Since qubits are shuttled forward and backward, the total number of shuttling steps corresponds to the number of quantum dots within the row. For upscaling, the critical consideration is the shuttling length against the qubit array size. In Fig. 2, we plot the required shuttling length (assuming a quantum dot pitch of 100nm) against the total number of qubits $N$ in the trilinear array. By considering shuttling lengths in the few-micron range, we obtain that our architecture can easily host a few thousands of qubits. For million-scale system sizes, the shuttling length is in the range of tens of microns, which is still comparable to the designed length in current shuttling-connected sparse architectures[37,38]. However, for systems beyond the million scale, the extensive shuttling length likely requires very high shuttling fidelity and qubit coherence, which could be very challenging to realize.

For extremely large systems, the single qubit rows in the top and bottom of the trilinear array could be replaced by multiple rows and, ultimately, by semi-2D arrays, as illustrated by the gray inset in Fig. 2 – at the expense of more complex interconnection layers (more back-end-of-line layers to sustain the semi-2D arrays). In the ultimate limit, this would result in $\sqrt{N}$ square 2D sub-arrays of $\sqrt{N}$ quantum dots each, which would then each require shuttling over their sub-array row length of $\sqrt{\sqrt{N}}$. so that the shuttling length can be reduced to a few microns – even for billion scale systems as shown by the gray line in Fig.

2. Interconnecting a semi-2D qubit array would require multi-layer wiring, and transferring qubits across the semi-2D array would require multiple operations[47,48], so this would need to be carefully analyzed in terms of potential tradeoffs vs shuttling length. As intermediate steps towards such semi-2D array, the 1D qubit arrays (single rows, aka arrays of size $1 \times \sqrt{N}$) at each side of the linear architecture can already be replaced by multiple-row qubit arrays ($M$ rows, aka arrays of size $M \times \sqrt{N}/M$) . This will require a limited increase in the number of interconnect wiring layers, while reducing the shuttling length to $(\sqrt{N}/M)$. Along the same lines, a few more 1D arrays can also be added to include more shuttling paths to enable more system operation protocols and avoid a single shuttling array resulting in bottlenecks. Of course, developing useful quantum computers with millions of interconnected physical qubits is a gargantuan task, and it may in practice take a large-scale effort, for which this architecture could provide a blueprint, to be augmented by dedicated and significant developments on materials, devices and the wiring layers themselves for the detailed implementation.

## Wiring interconnectivity and 3D integration

Small qubit arrays allow separate wiring from room temperature electronics to each qubit to demonstrate proof-of-principle quantum information processing[53,54]. For large qubit arrays, it is impractical to connect each gate electrode to a separate analog circuit – meaning that certain levels of multiplexing at the qubit-to-analogue-control interface should be implemented to address the interconnect bottleneck[32]. The proposed trilinear array architecture, in view of its accommodation of reduced effective pitches, lends itself well to a natural implementation that brings the control electronics to the qubit chip without room-temperature wiring. Our proposed approach encompasses two steps. First, we propose to contact each quantum dot gate through 3D integration, and second, to address those gates via multiplexed gate control involving cryoCMOS switch circuits.

In Fig. 3a, we show the qubit wiring scheme with the proposed 3D integration solution. The linear qubit architecture permits the fanout of all quantum dot gates to take place on a single chip plane, and the large space outside the linear array grants opportunities for contacting the qubits by staggered contact points with very relaxed pitches. The contact points can therefore easily reach an effective pitch in the micron range, unlike the 100 nm or below linear quantum dot gate pitch, which is well compatible with existing 3D integration techniques, offering in turn space for separate CMOS control of each quantum dot. Rather than cointegrating the qubits and cryoCMOS control electronics on the same chip, here we separate the qubit chip and the cryoCMOS chip. The quantum dot gate wiring is routed to through-silicon-vias (TSVs) in the qubit chip, and then directly 3D-bonded to the cryoCMOS chip. Such separation of the two modules allows for their individual optimization and, in addition, offers possibilities for thermal isolation between the two chips – the latter is important in view of the expected thermal load from large scale cryoCMOS chips and the need to keep electron temperatures of the quantum dot chips well below 1K. Alternatively, a flip-chip method can be used on the qubit chip to eliminate the TSV process[55], and an interposer chip or a redistribution layer can be added depending on the integration requirements[56].

In Fig. 3b, we illustrate the multiplexing scheme with cryoCMOS switch circuits. For each quantum dot gate, we split the voltage input based on the signal bandwidth and functionality. We use a low frequency (quasi-static DC) input for biasing static quantum dot working points, while a high frequency input (here referred to as AC, in contrast to the quasi-static DC control) serves for applying pulses for fast control and readout. These two inputs are ultimately combined by a bias-tee circuit and the sum of the DC and AC signals is applied to each quantum dot gate. Due to nonuniformity among different quantum dot sites[8,35], different DC biases may be required for the quantum dot plunger gates to reach the last-electron occupation in different quantum dots. Similarly, different barrier gates may need different DC biases to tune different dots into the same tunnel coupling range[57]. For these reasons, providing multiplexing capabilities for individual gate tuning is highly desired. We can enable this by using a sampling circuit, where charges stored on a capacitor maintain the voltage even after the input has been disconnected by opening the transistor switch[35]. This type of floating gate control is very similar to the storage mode in DRAM, but with the crucial difference that long holding times have been demonstrated experimentally at cryogenic temperatures due to the low-leakage rate of cryogenic CMOS – potentially allowing for very high multiplexing factors and low charge refresh rates[58,59].

For AC multiplexing, experimental demonstrations of time-domain operations have been reported with both spin and superconducting qubits[59,60]. Here, we take a step further and discuss the possibility of parallel operations with a limited set of AC inputs. Importantly, as we compensate for the nonuniformities among qubits by using the DC floating gates introduced above, one might be able to use a limited, finite set of distinct AC pulses for all qubits – down to potentially a single, common AC pulse control level with sufficient DC compensation and uniformity. In this ideal case, a finite, fixed set of AC pulses, for exchange control, shuttling, readout, would be needed regardless of the qubit number. Furthermore, we propose to employ an RF switch matrix that simultaneously routes the same AC pulses to different quantum dots, and ultimately allows for parallel operations of different qubits. The inputs of the RF switch matrix consist of its digital switch controls, as well as the fixed set of AC pulses for qubit operation and shuttling, and their total number scales sub-linearly with the number of qubits. A detailed example of qubit operation – with the multiplexing scheme is discussed in Supplementary Information S2.

## Practical operation considerations for a larger spin qubit array

When building larger quantum dot spin qubit arrays, another important challenge to tackle, besides the interconnect challenges discussed above, is the required high yield of quantum dots[26]. Disorder at the atomic scale could impact the confinement potential at multiple sites in the array – such that some quantum dots cannot be defined or the tunnel coupling between some dots cannot be controlled[57]. Moreover, in linear arrays that utilize shuttling, a single point of excessive disorder could break one system into two. In Fig. 4a, we show an example where one defective quantum dot site forbids the shuttling of qubits between its two sides – severely restricting the connectivity between the qubits on the left and on the right.

To mitigate this problem, the quantum dots surrounding the defective site can be reconfigured for shuttling at the cost of sacrificing a few discrete quantum dots to preserve the operation of the entire array, as shown in Fig. 4b. This reconfiguration is possible thanks to the individual wiring of each quantum dot, where the cryoCMOS circuits change the dot DC bias with the sampler and the AC pulses with the RF switch matrix. For larger qubit arrays, sacrificing a few qubits in this way will not strongly influence the overall performance of the array – as shuttling is preserved and the original connectivity can be achieved via only slightly longer shuttling distances for certain qubits. However, if there are multiple defective sites that happen to cut across the array, the above solution will not work. On the architectural level, multiple shuttling arrays or semi-2D qubit arrays, as shown in Fig. 2, can reduce the chance for such a scenario. Also, by considering advancements on the device level, high quantum dot yield rates can be achieved with advanced manufacturing, and further improvement is required for larger systems[26].

The connectivity between qubits is critical for efficient quantum algorithms and quantum error correction codes[61]. For example, nearest neighbor connectivity is essential for implementation of the surface code in 2D qubit grids. As discussed previously (Fig. 2), an $N$ qubit 2D grid can be mapped onto our trilinear architecture where the nearest neighbor connectivity can be preserved with a shuttling overhead of $\sqrt{N}$. Here, we highlight that higher orders of connectivity, between any two qubits in neighboring rows or columns, can be also achieved, with shuttling overheads of $3\sqrt{N}$ or less, as represented by the light blue lines in the upper 2D grid in Fig. 4c. One such example is highlighted by the purple line in the 2D grid, and its corresponding shuttling path is denoted in purple in the trilinear array below in Fig. 4d. Moreover, shuttling overheads of $2\sqrt{N}$ can also enable the connectivity between the qubits at the two opposite vertical edges of the 2D grid as illustrated by the green lines in Fig. 4d, which is topologically equivalent to folding the 2D grid into a tube. For a longer trilinear array, the two ends of the array could also be connected by arranging the array as a loop while still having enough space for wiring interconnects. The loop could then turn the tube into a donut, as shown in Supplementary Information S3. Such effective donut connectivity would satisfy topological requirements for forms of error correction that are otherwise out of reach in finite 2D arrays[33]. Finally, we emphasize that the trilinear architecture in principle allows for even higher orders of connectivity than those considered in Fig. 4c and d. Although, a detailed discussion of possible algorithms and error correction codes is beyond the scope of this work, we point out that such architectural capabilities could enable novel and more efficient quantum computation and error correction protocols.

## Conclusion and discussions

Qubit interconnect and wiring are arguably the central upscaling challenge for most quantum computing platforms. The small size of semiconductor quantum dot spin qubits offers great scalability on the qubit level with qubit pitches of around 100nm[62]. However, it raises many challenges for the 2D scaling of the interconnect layer, due to difficulties of wiring each quantum dot in a densely packed 2D array. In this work, we have proposed a scalable solution in the form of a trilinear quantum dot array in which the 2D connectivity is realized by qubit shuttling. This architecture allows for switch-based cryoCMOS modules for channel multiplexing and parallel qubit operation. Both the array structure and the control modules are compatible with the existing semiconductor and 3D integration technologies. The key requirements for fault-tolerant quantum computation in our trilinear architecture have so far been demonstrated in small scale systems. These include high quantum dot yield rate[26], single and two qubit gates beyond the fault-tolerant threshold[63], coherent shuttling[18] and operation of small-scale 1D arrays[64]. Next steps towards reaching such performance also in large qubit arrays would involve improving the uniformity of qubit devices by utilizing advanced semiconductor technology[25,26,65]. In the meanwhile, trilinear quantum dot arrays with intermediate qubit numbers could be realized to examine the architecture performance on the system-level.

The cost of the relatively simple layout of our trilinear architecture is the overhead in the qubit operation coming from qubit shuttling. Recent experiments have shown fast, high fidelity shuttling in the 10-micron range[18], which would correspond to the shuttling length required for a million-qubit system size in the architecture presented here. By using the proposed cryoCMOS control, different gate voltages could be simultaneously optimized to further improve the qubit performance at different sites. Moreover, shuttling can be used to extend the qubit connectivity beyond the nearest neighbors, which opens hardware possibilities for more efficient error correction schemes and algorithms[66].

Finally, this work has focused on the general layout of the trilinear architecture, and some more specific aspects of the functional qubit periphery have been omitted for clarity, as they scale linearly and are readily implementable in the proposed scheme. For completeness, we include in Supplementary Information S1 a detailed example of how the trilinear array architecture can be equipped with global qubit control, nano-magnet based qubit addressability, and single electron transistors for qubit readout. Importantly, the additions proposed there are fully compatible with existing CMOS technologies.

Nonetheless, we remark that all these different components bring specific requirements, and that building a full stack qubit architecture would need to be addressed as a system-level problem. We expect that the considered advantages of the trilinear qubit architecture motivate system-level developments in the design, fabrication, and operation of future quantum processors.

## Data availability

The data that support the findings of this study are available from the authors upon reasonable request

## Acknowledgements

The authors acknowledge financial support from European Union's Horizon 2020 Research and Innovation Program under grant agreement No 951852 (QLSI). This work was performed as part of IMEC's Industrial Affiliation Program (IIAP) on Quantum Computing. V.L., W. D. R., and K. D. G. acknowledge the support in part from the Excellence of Science (EOS) programme (FWO and F.R.S.-FNRS) through grant EOS G0H1122N EOS 40007526 CHEQS.

## Author contributions

R.L. and C.G. developed the physical qubit chip layout. S.K and S.B verified qubit and 3D integration schemes. G.S. analyzed prospects for quantum error correction. R.L. conceived the project with support from all authors. R.L., M.M., D.W., and K.D.G. supervised the project. R.L., V.L., and K.D.G. wrote the manuscript with input from all authors.

## Competing interests

R.L. C.G. S.K. G.S. B.R. are inventors on patent application related to this work, filed by IMEC (application no. EP 24166659.3; filing date, 27 March 2024). Additionally, R.L. C.G. S.K. G.S. D.W. K.D.G. are inventors on another patent application related to this work, also field by IMEC (application no. EP 25163893.8; filing date, 14 March 2025). The other authors declare that they have no competing interests.

# Figures

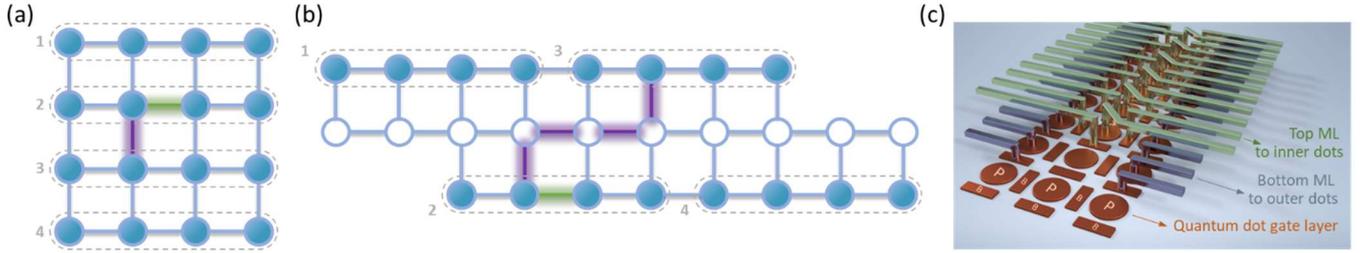

**Fig. 1 Design of the linear quantum dot architecture. (a)** Schematic representation of the 2D quantum dot array with square lattice and nearest neighbor connectivity. **(b)** The tri-linear array with equivalent connectivity as (a). The upper (lower) 1D array in the tri-linear architecture are equivalent to row 1 and 3 (2 and 4) of the 2D array as in (a). The middle of the tri-linear array consists of empty quantum dots for qubit shuttling. From a qubit connectivity perspective, interactions within the same row maintain the same operation scheme for the 2D and the tri-linear array, as noted by the green shaded lines connecting two qubit sites in (a) and (b) as an example. Interactions along the vertical directions between different rows require more steps on the tri-linear array than the 2D array, which needs moving the qubit, say, from the top 1D array to the middle and shuttle to the targeting site above the bottom array for two qubit operation, and then shuttle back to the original site. An example operation in the tri-linear array, and the equivalent one in the 2D array, are shown by the purple shaded lines in (a) and (b). **(c)** Three-dimensional model of the array gate structure. Two metal layers (MLs) allow individual connection to the plunger (P) and barrier (B) gates of each quantum dots. Some metal layers at the bottom left of the figure are removed for clarification.

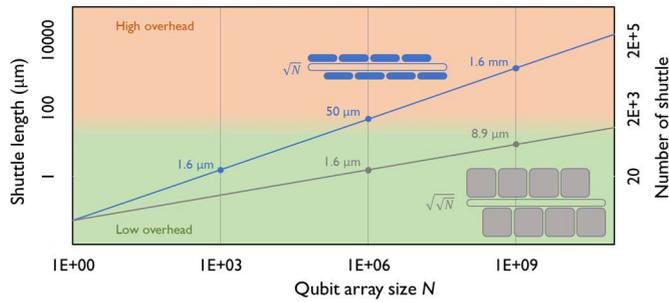

**Fig. 2 Shuttle length vs. different qubit array sizes.** In comparison to a 2D lattice, the linear qubit structure requires extra shuttling overhead for two qubit operations along the vertical directions. The blue line shows the shuttle length in the tri-linear array, where the number of shuttles needed is $\sqrt{N}$ – scaling with the number of rows in a 2D grid with N qubits; each of those shuttles also requires on the order of sqrt(N) empty quantum dots to be able to connect neighboring rows in the 2D grid. For thousands of qubits, the shuttle length is in the few microns scale, which is well within the low overhead regime. For million scale qubit systems, the shuttling length reaches the tens of micron range, which is still within the reasonable overhead region but could require excessively high shuttling fidelity and speed. For very large-scale systems, the 1D qubit array at the top and bottom of the tri-linear structure can be replaced by semi-2D arrays, as illustrated by the gray inset at the bottom right of the plot. Both the number of shuttles as well as the length of each shuttle then further reduces to $\sqrt{\sqrt{N}}$, and the shuttle length can be in the range of few microns even for billion scale systems. A quantum dot pitch of 100 nm is used for the shuttle length calculation.

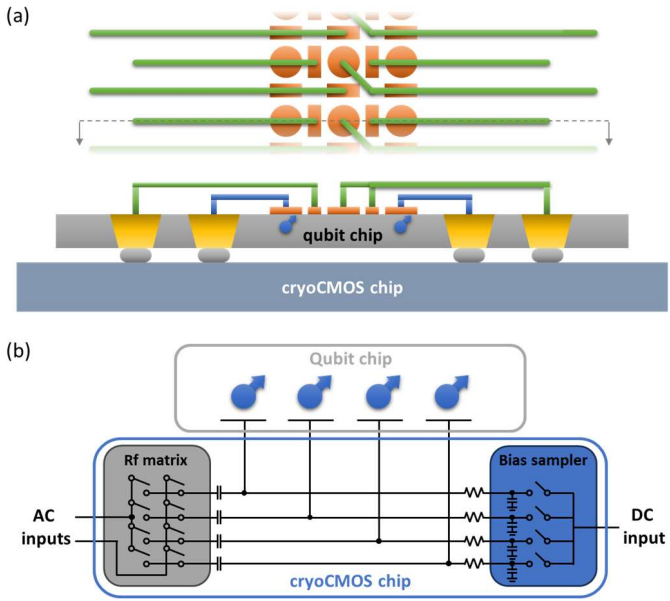

**Fig. 3 3D integrated qubit and cryoCMOS chips for interconnectivity. (a)** Schematic of the 3D integrated qubit system. Outside the tri-linear array, there is ample space to interconnect each qubit gate and ensure that realistic wiring pitches can be used. By interleaving the wiring ending points, each quantum dot gate can be routed to interconnect structures having dimensions much larger than the pitch of the quantum dots. In this schematic, we use TSVs (through-silicon-vias) to bring the interconnect to the back of the qubit chip. A similar approach with flip-chip bonding to interposers or control chips on the front side of the chip can also be considered (not shown). The qubit chip is subsequently 3D integrated to a cryoCMOS chip for qubit control. **(b)** The scheme for scalable interconnection to each quantum dot gate with the cryoCMOS chip. Each gate needs two inputs: DC to set the gate bias point and AC for shuttling and qubit operations. For DC, different gate biases are stored with a sampling circuit so that a single input can provide DC bias voltages for multiple gates in a time-based multiplexing manner. For AC, an RF switch matrix can be used to route different sets of AC pulses to different sets of gates simultaneously, allowing for parallel qubit operations with a limited number of AC inputs. The exact number of AC inputs in this scheme depends on the degree of homogeneity achieved after tuning each dot carefully using the DC biasing scheme.

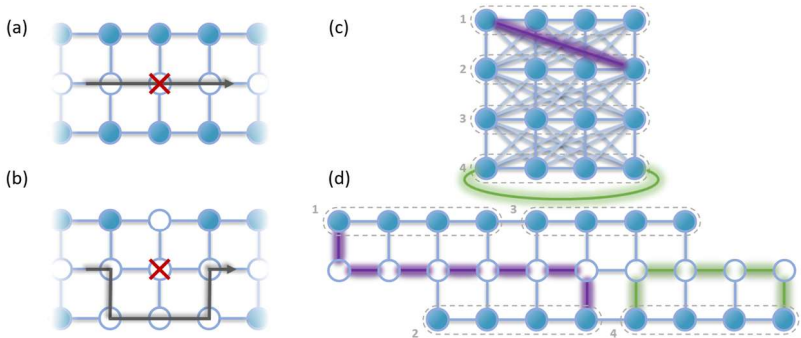

**Fig. 4 Practical qubit array operation considerations. (a) and (b)** Scheme to avoid defect quantum dots in the tri-linear array. The qubit quantum dots around the defective site can be configured as empty shuttling quantum dots to fix the connectivity. **(c) and (d)** Schemes for higher order of qubit connectivity. (c) shows a 2D qubit array with square lattice that has row-wise all to all connectivity (connectivity represented by the light blue lines connecting different qubits). In the tri-linear array having N qubits as shown by (d), such a connectivity can be achieved with a shuttling overhead of $3\sqrt{N}$. The purple and green shaded line shows the example of connecting qubits beyond the nearest neighbour at different rows and within the same row, respectively.

# Supplementary information
# A trilinear quantum dot architecture for semiconductor spin qubits

R. Li*[1], V. Levajac[1,3], C. Godfrin[1], S. Kubicek[1], G. Simion[1], B. Raes[1], S. Beyne[1], I. Fattal[1,2], A. Loenders[1,2], W. De Roeck[3], M. Mongillo[1], D. Wan[1], K. De Greve[1, 2]

[1] IMEC, Leuven, Belgium
[2] Department of Electrical Engineering, KU Leuven, Leuven, Belgium
[3] Department of Physics, KU Leuven, Leuven, Belgium

## Section 1, Sample implementation of a fully functional 1024 qubit chip based on the trilinear architecture

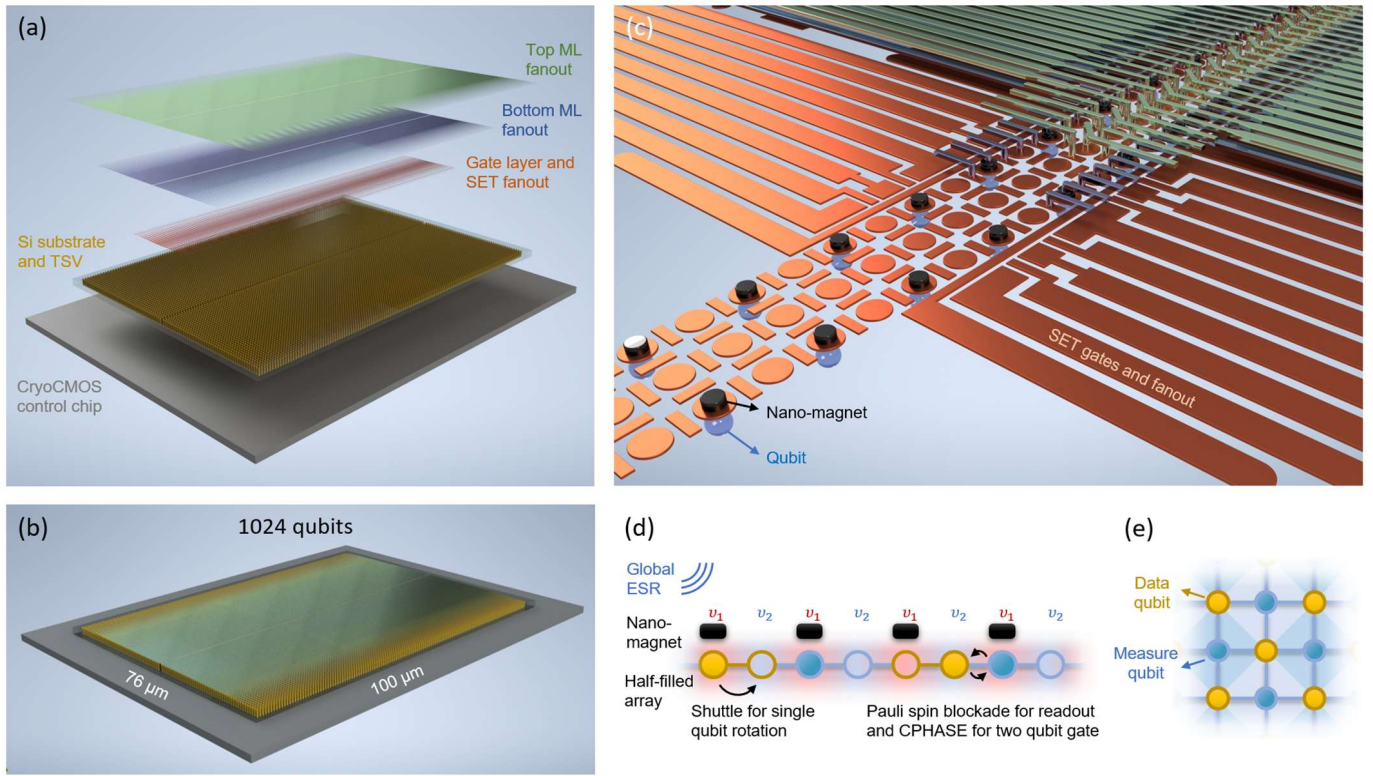

**Supplementary Fig. 1 An example layout of a fully functional 1024-qubit chip. (a)** The layered breakdown of the qubit chip, including, from bottom to top, (i) the cryoCMOS control chip, (ii) the qubit-hosting silicon substrate and TSVs which connect the control wiring from the cryoCMOS to each quantum dot and SET gate, (iii) the quantum dot and SET gates and fanout connection to TSVs, (iv) the bottom metal layer and fanout, and (v) the top metal layer and fanout. **(b)** A 3D schematic of the qubit chip. For 1024 qubits, the qubit array and gate fanout size is ~ 76 μm × 100 μm. **(c)** A zoomed-in image of the trilinear array. SETs are attached to both sides of the trilinear array for charge sensing. On every other quantum dot in the qubit array, a nanomagnet is placed on top of the plunger gate, which is also used as electrical connection to that plunger gate. **(d)** The operation scheme of the qubit chip. **(e)** The corresponding 2D grid where the blue and yellow colored qubits can be used as data and measure qubits for surface code implementations.

Here we show an example implementation of a fully functional qubit chip containing 1024 qubits. The 3D schematics are shown in Fig. S1 a-c. Most of the chip area is used for wiring fanout. The length of the vertical wiring fanout to the trilinear array is set by the size of the TSVs, which have a pitch of 0.8 μm as used in the schematic plots.

Qubit operations are based on a half-filled array[1] with periodic local magnetic fields as shown in Fig. S1d. Applying an out-of-plane static magnetic field defines the spin quantization axis. A nanomagnet is placed on top of every other plunger gate in the 1D quantum dot (QD) array. This arrangement of nanomagnets generates a periodic magnetic field oscillation along the array between resonance frequencies $v_1$ and $v_2$, giving rise to two different qubit frequencies depending on if a QD is under a nanomagnet or not. We propose a realization where QD arrays are half-filled with qubits, with QDs under nanomagnets ($v_1$) having qubits and QDs without nanomagnets ($v_2$) kept empty. When a $v_2$-resonant global ESR $v_2$ is applied, a qubit can be shuttled to its neighbouring nanomagnet-less QD where it will be resonant with the ESR and exhibit a single qubit gate rotation. In this way, qubits can be selectively targeted to either idle in QDs or shuttle to empty QDs to be operated on, thereby allowing addressable single qubit gate operations with a global ESR. The frequency difference between $v_1$ and $v_2$ also allows two-qubit gate operations via CPHASE and Pauli spin blockade-based readout. Along a 1D array, a qubit could alternatively serve as data and measurement qubit. Extrapolating to 2D, as shown in Fig. S1e, would allow for surface code operations[2].

It is worth noting that shutting in the middle quantum dot array could cause additional Z rotations. Nonetheless, this rotation can be calibrated and compensated (e.g. by virtual Z gates).

## Section 2, Wiring interconnect and multiplexing scheme with cryoCMOS

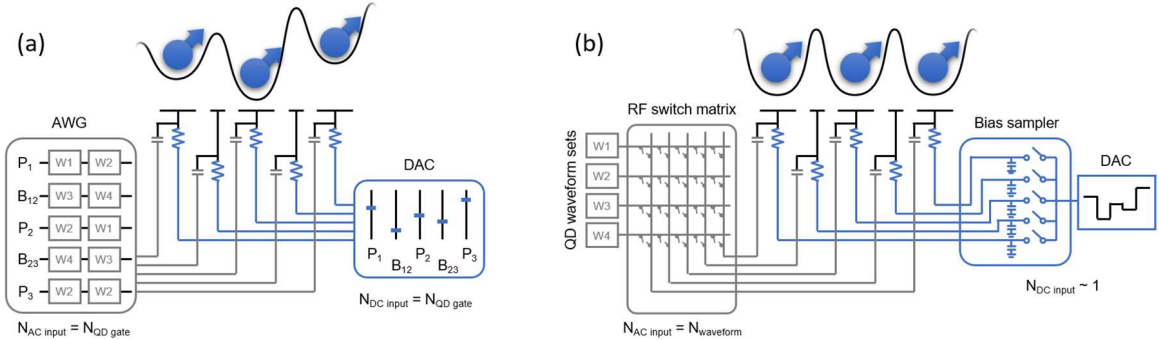

**Supplementary Fig. 2 Qubit wiring and control schemes. (a)** The conventional wiring scheme, where separated DAC and AWG channels are needed for each quantum dot gate. **(b)** The multiplexing scheme with a cryoCMOS chip, where a fixed number of AC and DC inputs are needed to control a large qubit array.

To optimize qubit operation fidelities, typically each gate needs dedicated AC and DC controls, as shown in Fig. S2a. To achieve a scalable wiring and interconnect scheme, which should scale sub-linearly with the number of qubits[3], a cryoCMOS circuits chip is proposed, as shown in Fig. S2b. The cryoCMOS chip contains a bias sampler and an RF switch matrix. For each dot gate, the bias sampler uses a switch to couple an external DAC for voltage biasing, and charge storing capacitors store the voltage after opening the switch and decoupling from the DAC. This scheme is known as the floating gate scheme[4]. With low charge leakage rate on the capacitor and the quantum dot gate (more than one hour holding time)[5], a single DC input could support hundreds of quantum dot gates by sequentially changing the DAC voltage to the corresponding quantum dot gate voltage once per second, for example, and then refreshing the corresponding floating gate capacitor.

For AC input, we assume that once the static biasing point is corrected by the floating gate DC bias, the same AC pulse can be used across all quantum dot gates. For typical qubit operations, only a limited set of waveforms are required, including qubit shuttling in the middle dot array (e.g. 4 waveforms are needed for conveyor-mode shuttling[6]), single qubit hopping in the outer dot arrays, two qubit gates, readouts, and each compensation gate pulses. These waveforms can be periodically generated with external AC inputs and selectively distributed to different quantum dot gates (and simultaneously for parallel qubit operation) through the RF switch matrix. In this case, even with increased qubit numbers, the number of AC inputs is fixed (as the total required waveforms is not related to the qubit array size).

The above discussion assumes that the quantum dots have good uniformity for shared AC control (or good capacitance matrix uniformity to be more specific[7]). Meanwhile, if there are few outlier dots, more input channels can be included on the RF switch matrix to allows for dedicated AC control input for the outlier dots. If the uniformity of the dots is not sufficiently good, time-domain multiplexing, or row by row operation has to be used[8].

## Section 3, Sample loop connection scheme for the trilinear array

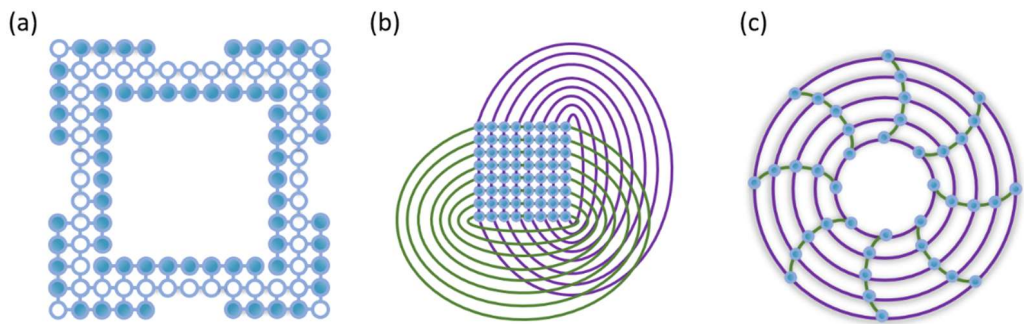

**Supplementary Fig. 3 Loop connection scheme for the trilinear array. (a)** The schematic layout of a trilinear array that is equivalent to an 8×8 2D array. The head and tail of the trilinear array are connected together, forming a loop. **(b)** The equivalent connection scheme of (a). The green lines show the connection within the same row by a shuttling length of $2\sqrt{N}$, same as the green line in main text Fig. 4c. The purple lines show the connection of column heads and tails through the loop of the connection of the trilinear array. **(c)** The donut equivalent connection scheme of (a) and (b), where the standard 2D array edge is eliminated. Note that qubits and connections at the backside of the donut are not shown.

Fig S3 shows the connection scheme of the trilinear array to form a donut to eliminate all edges on a square 2D array. Here we use an 8×8 2D array for illustration purposes. Larger arrays could allow for even easier layout as well as more space inside the loop for fanout. Moreover, by curving the trilinear array or meandering the array, the chip areas can be more effectively utilized for large systems, as the straight 1D layout can end up very long.

## Supplementary references